# Deep learning-based detection of intravenous contrast in computed tomography scans

Zezhong Ye[1,2], Jack M. Qian[1,2], Ahmed Hosny[1,2], Roman Zeleznik[1,2], Deborah Plana[1,3], Jirapat Likitlersuang[1,2], Zhongyi Zhang[1,2], Raymond H. Mak[1,2], Hugo J. W. L. Aerts[1,2,4,5], Benjamin H. Kann[1,2]*

1. Artificial Intelligence in Medicine (AIM) Program, Mass General Brigham, Harvard Medical School, Boston, MA, USA
2. Department of Radiation Oncology, Dana-Farber Cancer Institute and Brigham and Women's Hospital, Harvard Medical School, Boston, MA, USA
3. Harvard-MIT Division of Health Sciences & Technology, Cambridge, MA, USA
4. Department of Radiology, Brigham and Women's Hospital, Dana-Farber Cancer Institute, Harvard Medical School, Boston, MA, USA
5. Radiology and Nuclear Medicine, CARIM & GROW, Maastricht University, Maastricht, the Netherlands

* Corresponding author

**Correspondence address to:**
Benjamin H. Kann, M.D.
Department of Radiation Oncology,
Dana-Farber Cancer Institute and Brigham and Women's Hospital,
Harvard Medical School, 75 Francis Street, Boston, MA 02115, MA, USA
Tel: +1 617-732-6310
Email: Benjamin_Kann@dfci.harvard.edu

**Summary statement**
We demonstrated that a CNN-based deep learning model accurately detects IV contrast enhancement in the CT scans from both head and neck and lung cancer patients efficiently and with near perfect performance. This pipeline can be implemented to assist data curation and quality control for a variety of clinical and research applications, saving significant time and manual effort and improving accuracy of contrast annotation in imaging studies.

**Key points**

1. We utilized 1,979 head and neck and chest CT scans from five institutions and one national clinical trial to develop and validate a deep learning model to detect IV contrast.
2. An EfficientNetB4-based deep learning model yielded AUC: 0.996 in the internal validation set (n = 216) and 1.0 in the external validation set (n = 595) for head and neck CT scans.
3. Utilizing transfer learning and fine-tuning, the model was retrained to detect IV contrast on chest CT scans with AUC: 1.0 for the internal validation set (n = 53) and AUC: 0.980 for the external validation set (n = 402).

**ABSTRACT**

**Purpose:** Identifying intravenous (IV) contrast use within CT scans is a key component of data curation for model development and testing. Currently, IV contrast is poorly documented in imaging metadata and necessitates manual correction and annotation by clinician experts, presenting a major barrier to imaging analyses and algorithm deployment. We sought to develop and validate a convolutional neural network (CNN)-based deep learning (DL) platform to identify IV contrast within CT scans.

**Methods:** For model development and evaluation, we used independent datasets of CT scans of head, neck (HN) and lung cancer patients, totaling 133,480 axial 2D scan slices from 1,979 CT scans manually annotated for contrast presence by clinical experts. Five different DL models were adopted and trained in HN training datasets for slice-level contrast detection. Model performances were evaluated on a hold-out set and on an independent validation set from another institution. DL models was then fine-tuned on chest CT data and externally validated on a separate chest CT dataset.

**Results:** Initial DICOM metadata tags for IV contrast were missing or erroneous in 1,496 scans (75.6%). The EfficientNetB4-based model showed the best overall detection performance. For HN scans, AUC was 0.996 in the internal validation set (n = 216) and 1.0 in the external validation set (n = 595). The fine-tuned model on chest CTs yielded an AUC: 1.0 for the internal validation set (n = 53), and AUC: 0.980 for the external validation set (n = 402).

**Conclusion:** The DL model could accurately detect IV contrast in both HN and chest CT scans with near-perfect performance.

**INTRODUCTION**

There is growing interest in artificial intelligence (AI) for medical imaging analysis and the potential to improve clinical care (1–3). However, there exist many challenges that have prevented large-scale clinical translation of the numerous algorithms in development(2). Among these challenges, perhaps the most critical is healthcare data quality and curation. Data curation and quality control for medical imaging analysis projects are extremely difficult, labor-intensive, and largely manual processes, involving scan-by-scan examination of the imaging data by clinical experts (4). Imaging data is generated from a variety of scanners and clinical protocols which can have heterogeneous physical parameters (i.e. pixel spacing, reconstruction kernel, energy) that must be characterized systematically, and sometimes normalized, prior to analysis (5). Characterization of these parameters relies on metadata tags from the Digital Imaging and Communications in Medicine (DICOM) standard, which was developed as a clinical protocol and not designed for downstream computational analysis (6). Furthermore, certain types of metadata are input manually by scan operators and are notoriously poorly documented and error prone (7–9). One such parameter is the administration of intravenous (IV) contrast agents (8,9).

IV contrast is used routinely to enhance the diagnostic yield of computed tomography (CT) scans (10). Unsurprisingly, presence or absence of IV contrast is known to have huge ramifications for computational imaging model performance, and it is essential to know the contrast status of the scans within a dataset for imaging analyses(11–14). Currently, the only reliable way to designate contrast status in a scan is via manual review by clinical experts, which is time-consuming and often impractical. With growing interest in using large datasets to better train computational models, there is urgency to develop tools that can auto-detect IV contrast with high fidelity. In the area of head and neck radiology, there has been a growing number of quantitative imaging studies utilizing AI, including segmentation (15), classification (11,12), radiogenomics (16), radiotherapy planning (17), and outcome prediction (18) that rely on curated training data with a knowledge of contrast presence or absence. Currently, studies may rely on (unreliable) metadata or manual classification of contrast status, and it often remains unclear how these processes are carried out in practice. Deep learning has demonstrated tremendous promise for medical imaging classification (19). Currently, there is no reliable automated contrast detection model for head and neck (HN) CT, and no deep learning-based models for contrast detection for CT of any anatomic region.

In this study, we hypothesized that a convolutional neural network (CNN)-based deep learning algorithm could reliably and accurately detect IV contrast in CT scans. We sought to develop an automated contrast detection platform to streamline data curation and quality control, improving the feasibility of large-scale quantitative imaging analyses. We anticipate such a tool could accelerate the development of clinically deployable AI and be useful for a variety of clinical and research applications.

## MATERIALS AND METHODS

### Study design and datasets

This study was conducted in accordance with the Declaration of Helsinki guidelines and following the Mass General Brigham (MGB) Institutional Review Board (IRB) approval. Waiver of consent was obtained from the MGB IRB prior to research initiation. The HN cancer dataset consists of four publicly available, de-identified patient cohorts, each downloaded and curated from The Cancer Imaging Archive (TCIA) (20), as follows: 1) PMH cohort (n = 558) (21); 2) CHUM cohort (n = 61) & CHUS cohort (n = 101) (22); 3) MDACC cohort (n = 603) (23). The lung cancer dataset includes two patient cohorts: 1) Harvard-RT cohort (n = 262); 2) RTOG-0617 cohort (n = 402)(24). These subsets represent all scans that passed initial quality control of DICOM metadata and were able to be successfully converted to 3D Nearly Raw Rasterized Data (NRRD) format. Scans without head and neck portions (n = 3) and whole-body scans (n = 6) were excluded for the analysis (Fig. 1) (Supplement).

| Patient Cohort (n = 1315) | Training (n = 504) | Internal Validation (n = 216) | External Validation (n = 595) |
|---|---|---|---|
| **Age** | | | |
| median (range) | 61 (34 – 90) | 59 (33 – 88) | 57 (24 – 91) |
| **Sex n (%)** | | | |
| Female | 109 (21.6%) | 46 (11.4%) | 83 (13.9%) |
| Male | 395 (78.4%) | 170 (78.6%) | 493 (82.9%) |
| Unspecified | 0 | 0 | 19 (3.2%) |
| **Primary Cancer Site n (%)** | | | |
| Oropharynx | 482 (95.6%) | 201 (93.1%) | 525 (88.2%) |
| Larynx/Hypopharynx/Nasopharynx | 18 (3.6%) | 14 (6.5%) | 15 (2.5%) |
| Oral cavity | 0 | 0 | 7 (1.2%) |
| Unknown/Other | 4 (0.8%) | 1 (0.5%) | 48 (8.1%) |
| **Clinical Stage n (%)** | | | |
| I | 8 (0.8%) | 6 (0.4%) | 6 (1.0%) |
| II | 36 (14.7%) | 18(17.1%) | 16 (2.7%) |
| III | 76(36.1%) | 29 (33.3%) | 78 (13.1%) |
| IV | 383 (29.6%) | 162(31.9%) | 476 (80.0%) |
| Unspecified | 0 | 0 | 19 (2.9%) |
| **HPV/p16 Status* n (%)** | | | |
| Negative | 111 (22.0%) | 39 (18.1%) | 13 (2.2%) |
| Positive | 250 (49.6%) | 122 (56.5%) | 242 (40.7%) |
| Unspecified | 143 (28.4%) | 55 (25.5%) | 340 (57.1%) |
| **IV Contrast via DICOM Metadata** | | | |
| Contrast | 99 (19.6%) | 48 (22.2%) | 388 (69.7%) |
| Non-Contrast | 0 | 0 | 63 (9.7%) |
| Missing | 405 (80.4%) | 168 (77.8%) | 196 (30.3%) |
| **IV Contrast via Expert Annotation** | | | |
| Contrast | 270 (53.6%) | 116 (53.7%) | 411 (69.1%) |
| Non-Contrast | 234 (46.4%) | 100 (46.3%) | 184 (30.9%) |

**Table 1.** Head and neck cancer patient characteristics. *Patients with non-oropharyngeal carcinoma who did not undergo HPV or p16 testing were coded as negative, given the very low incidence of HPV/p16 positive tumors in these disease sites.

| Patient Cohort (n = 664) | Harvard-RT Training (n = 209) | Harvard-RT Internal Validation (n = 53) | RTOG0617 External Validation (n = 402) |
|---|---|---|---|
| **Age** | | | |
| median (range) | 65 (33 – 84) | 65 (41 – 80) | 64 (37 – 83) |
| **Sex n (%)** | | | |
| Female | 111 (53.1%) | 28 (52.8%) | 155 (38.5%) |
| Male | 98 (46.9%) | 25 (47.2%) | 203 (55.5%) |
| Unspecified | 0 | 0 | 32 (6.0%) |
| **Clinical Stage n (%)** | | | |
| IIA | 8 (3.8%) | 1 (1.9%) | 0 |
| IIB | 13 (6.2%) | 0 | 0 |
| IIIA | 126 (60.3%) | 30 (56.6%) | 248 (61.7%) |
| IIIB | 62 (29.7%) | 22 (41.5%) | 130 (32.3%) |
| Unspecified | 0 | 0 | 24 (6.0%) |
| **Histologic Type (%)** | | | |
| Adenocarcinoma | 129 (62.0%) | 29 (56.9%) | 153 (38.1%) |
| Adenocarcinoma in situ | 1 (0.5%) | 0 | 0 |
| Adenosquamous carcinoma | 3 (1.4%) | 0 | 0 |
| Squamous cell carcinoma | 45 (21.6%) | 13 (25.5%) | 164 (40.8%) |
| Large cell neuroendocrine carcinoma | 24 (11.5%) | 7 (13.7%) | 0 |
| Non-small cell lung cancer NOS | 4 (1.9%) | 0 | 52 (12.9%) |
| Mixed NSCLC and SCLC | 1 (0.5%) | 1 (2.0%) | 0 |
| Large cell undifferentiated | 0 | 0 | 9 (2.2%) |
| Unspecified | 1 (0.5%) | 1 (2.0%) | 24 (6.0%) |
| **IV Contrast via DICOM Metadata** | | | |
| Contrast | NA | NA | NA |
| Non-Contrast | NA | NA | NA |
| Missing | 209 (100%) | 53 (100%) | 402 (100%) |
| **IV Contrast via Expert Annotation** | | | |
| Contrast | 79 (37.8%) | 20 (37.7%) | 98 (24.4%) |
| Non-Contrast | 130 (62.2%) | 33 (62.3%) | 304 (75.6%) |

**Table 2.** Lung cancer patient characteristics. NSCLC = non-small cell lung cancer. SCLC = small cell lung cancer.

**CT scan characteristics and image acquisition**

CT scans were performed on 12 different CT scanner models from multiple institutions. Scanner specifications and imaging protocol details can be found in Tables S2-S7. CT scans were diagnostic quality, using 120-140 kVp energy, slice thickness 1-5 mm, and pixel spacing 0.3-2.7 mm. Presence of iodinated intravenous contrast from metadata was captured via DICOM tag (0018, 0010), "Contrast/Bolus Agent". Pretreatment CT scans for the patients were de-identified and exported in their entirety as decompressed DICOM files.

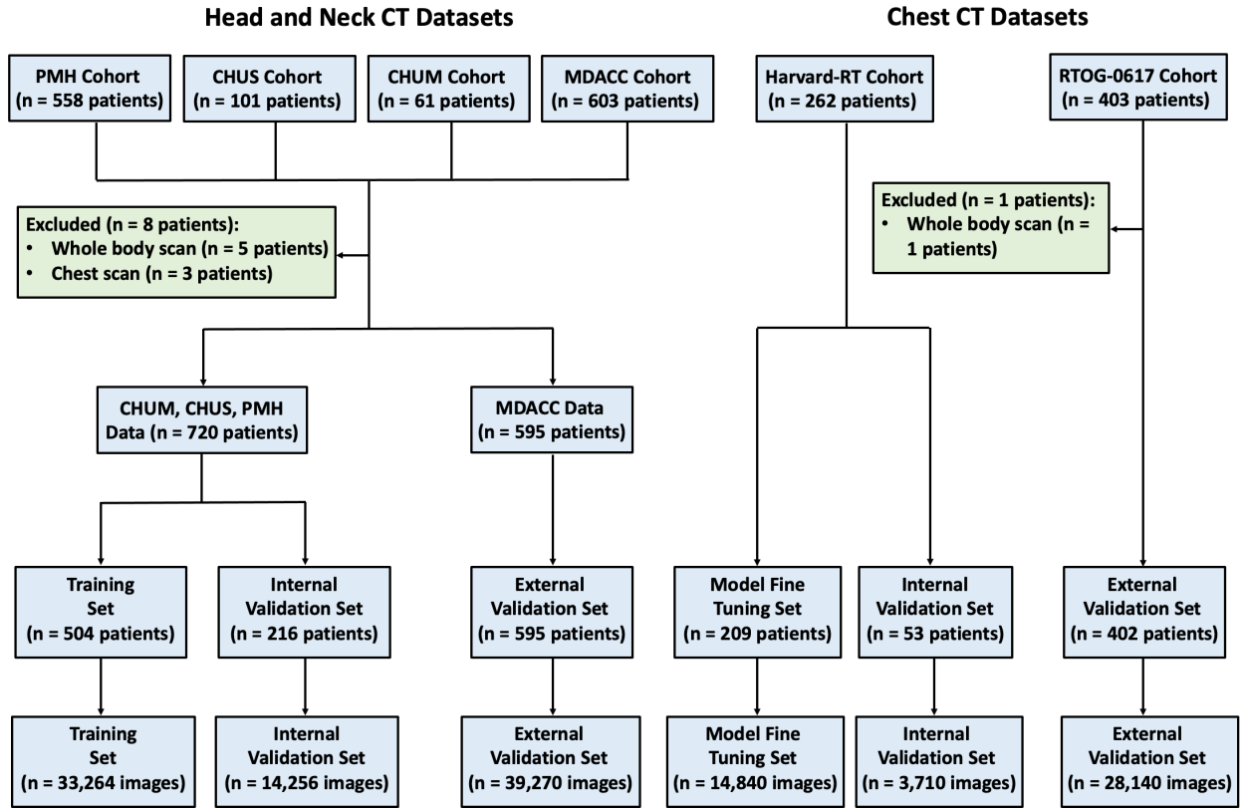

**Figure 1.** Consort diagrams for training and validation cohorts. PMH: subset of radiotherapy planning CT scans retrieved for all non-metastatic p16-positive Oropharynx cancer (OPC) patients treated with radiotherapy or chemoradiotherapy at Princess Margaret Cancer Centre between 2005 and 2010. CHUS & CHUM: subsets from FDG-PET/CT and radiotherapy planning CT imaging data of 298 head and neck cancer patients from four different institutions in Québec between 2006 and 2014. MDACC: subset from 627 head and neck squamous cell carcinoma (HNSCC) patients treated at MD Anderson Cancer Center. Harvard-RT: subset of head and neck cancer patients treated with radiation therapy at Dana-Farber Cancer Institute. RTOG-0617: subset of the study of high-dose or standard-dose radiation therapy and chemotherapy with or without cetuximab in treating patients with newly diagnosed stage III non-small cell lung cancer that cannot be removed by surgery.

**Image review and annotations**

All of the CT images were manually reviewed and annotated at the image axial slice- and scan-level for IV contrast presence by a radiation oncologist (J.M.Q.) with four years of clinical experience, and then were further reviewed by a board-certified radiation oncologist with seven years of clinical experience (B.H.K.) to confirm. Faint contrast, metal artifacts, and different scan fields of view (FOVs) were also documented and annotated.

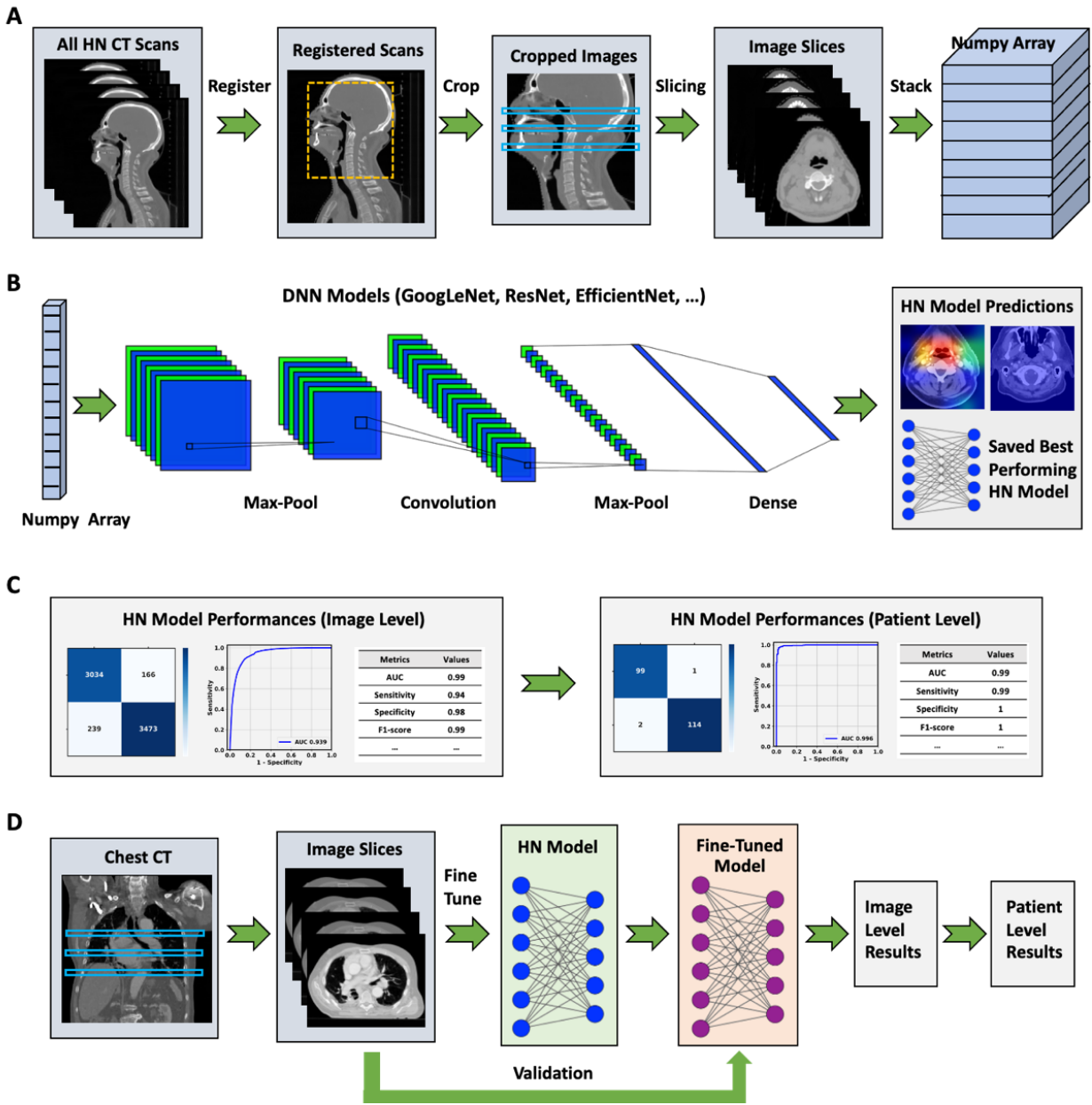

**Figure 2.** Workflow of deep neural networks (DNN) for contrast detections. (A) All the head and neck (HN) cancer CT scans were first co-registered to each other. The scans were then cropped to include only head and neck portions and exclude most background areas. 2-dimensional (2D) image slices were extracted from each scan and stacked together before converted to numpy arrays. (B) Numpy arrays with corresponding labels of each image slice were feed into DNNs for model development and validation. We tested multiple published 2D DNNs including

ResNet101V2, EfficientNetB4, InceptionV3 as well as simple convolutional neural network (CNN). The models and prediction results were saved. (C) Image-level model performances were evaluated directly from model predictions. Patient-level model performances were then calculated by averaging the probability scores of each image slices from each patient. (D) Chest CT scans went through the same imaging preprocessing before input for training. We used a portion of lung images to fine tune the saved models from HN datasets and applied other portions of lung images to validate the model performances on both image levels and patient levels.

**CT image preprocessing**

The CT images were first converted from DICOM format to NRRD format via rasterization packages utilizing SimpleITK (25) and plastimatch (https://plastimatch.org). All scans were resampled to 1×1×3 mm$^3$ pixel sizes using linear interpolation via SimpleITK. After interpolation, the CT scans were co-registered using a linear registration step with SimpleITK. The registration was performed separately for the HN cancer cohorts and the lung cancer cohorts. To reduce memory usage and computational load, we further applied a cropping step on three dimensions for each of the scans (Fig. 2A) (Supplement).

**Deep neural network architectures**

We investigated one simple CNN model (Fig. S1) and three representative published deep CNN models that have been top performers in classifications of large imaging datasets: ResNet101V2 model (26), InceptionV3 model (27) and EfficientNetB4 model (28). In brief, our simple CNN model was constructed with 4 convolutional layers and 2 fully connected layers. We also investigated using a transfer learning approach based on ResNet101V2 with pre-training weights on ImageNet (29). Model architecture details are found in the Supplement.

**CNN model implementation, training and validation**

After data preprocessing, HN scans from three patient cohorts (PMH, CHUM and CHUS) were shuffled and randomly split into 70:30 for model training (n = 504 patients, 33,264 images) and internal validation (n = 216 patients, 14,256 images). The data partition was stratified by IV contrast enhancement to keep equivalent ratios of contrast scans and non-contrast scans in each set. Additional HN scans from MDACC cohort (n = 595 patients, 39,270 images) were used for independent external validation on the model. We first trained, tuned, and tested the model on the image level. Each CNN model was trained for up to 100 epochs on the training dataset and validated on the validation set. We constructed and trained the CNN models using TensorFlow 2.0 frameworks in Python v3.8. The algorithm was trained on a Titan RTX graphics processing unit (Nvidia; Santa Clara, CA). See Supplement for further details.

**Model fine tuning and validation on chest CT scans**

To determine if the model largely based on HN CT scans could generalize to chest CT, we used 80% of the Harvard-RT chest CT dataset (n = 209 patients, 14,840 images) to fine-tune the HN model and used the remaining data (n = 53 patients, 3,710 images) for internal validation. Additional scans from RTOG-0617 dataset (n = 402 patients, 28,140 images) were used for external validation for the model. For fine-tuning, we allowed all network layers to retrain using a low learning rate ($10^{-5}$) using 10 epochs.

**Model performances and statistical analysis**

We trained and tuned the model based on the accuracy and loss at the image level. After development of the model, we locked the model and evaluated model performances at both the image level and patient level on the external validation set. We primarily evaluated the model diagnostic performance at the patient level using the patient probability score calculated by averaging the probability scores of all the images of each scan (eq 1).

$$\text{patient probability score} = \frac{\sum \text{image probability scores}}{\text{number of image slices for a given patient}} \quad \text{(eq 1)}$$

Confusion matrices were calculated to demonstrate the agreement between model predictions and gold standard labels. We performed receiver operating characteristics (ROC) analysis and calculated area under the curve (AUC) to assess model discrimination on IV contrast. Sensitivity and specificity values were calculated using the optimal cut off point with Youden Index. Because the dataset included imbalanced classes, the precision-recall curves and $F_1$-scores were calculated to provide complementary information to the ROC curve. 95% confidence intervals were calculated based on results over 10,000 bootstrapped iterations (30). Statistical metrics and curves were calculated using Scikit-learn (31) packages in Python v3.8. The source code and model can be found at: https://github.com/xmuyzz/DeepContrast.

**Gradient-weighted class activation mapping (Grad-CAM) analysis**

We selected the best overall performing models and calculated activation heat maps using Gradient-weighted class activation mapping (Grad-CAM) (32). We used the gradient information flowing into the last convolutional layer of CNN to assign values of importance to each element in the feature map for Grad-CAM (33). Finally, Grad-CAM maps were overlaid on corresponding CT images to provide interpretable visual explanations to model predictions.

**RESULTS**

**Patient and CT scan characteristics**

*Head and neck CT scans*

The total patient cohort consisted of 1,315 HN cancer patients (Table 1). Manual contrast annotation required 7.6 hours of clinician time for all of the HN scans (n = 1,315) and yielded 798 (60.7%) scans to be contrast-enhanced. There were 491 scans (67.8%) documented with artifact (minimal: 30; moderate: 129; severe: 149; very severe: 31) among 724 patients coming from PMH, CHUM, and CHUS datasets (Table S1). Additionally, 28 scans (2%) were determined to have "faint" contrast enhancement. In the DICOM metadata, there were 54 different types of IV contrast agent bolus tags (Table S9). DICOM metadata for contrast information was missing or erroneous in 808 scans (61.4%).

*Chest CT scans*

The lung cancer patient cohort consisted of 664 chest scans (Table 2). Manual review of the scans from RTOG-0617 dataset (n = 402) required 3.7 hours of clinician time for the chest scans and yielded 98 (24.4%) of scans to be contrast-enhanced. There were 10 (2.5%) scans determined to have "faint" enhancement. DICOM metadata for IV contrast was missing for all the chest CT scans.

**Model performance on head and neck CT scans**

All five models tested (simple CNN, ResNet101V2, EfficientNetB4, InceptionV3, and transfer learning; see Methods) yielded excellent results with patient-level AUCs of >0.98 and $F_1$-score of >0.96 on internal hold-out validations sets and external validation sets at the patient level (except for the simple CNN) (Table S10&S11). The EfficientNetB4 (Table 3 & Fig. 3) yielded the best overall performance (AUC: 0.996; F1-score: 0.991). On evaluation of performance metrics on the external validation set, the EfficientNetB4 yielded perfect patient-level classification performance with AUC: 1.0 (95% CI: 1.0-1.0), sensitivity: 100% (95% CI: 100%-100%), specificity: 100% (95% CI: 100%-100%) and $F_1$-score: 1.0 for the patient-level prediction (Table 3 & Fig. 3B). Additionally, the EfficientNetB4 model demonstrated AUC: 0.988 (95%CI: 0.988-0.988), sensitivity: 95.9% (95.9%-96.0%), specificity: 96.6% (96.6%-96.7%), and $F_1$-score: 0.964 on

image-level prediction. The EfficientNetB4 confusion matrices for the internal validation set (Fig. 4A) and external validation set (Fig. 4B) both demonstrated excellent agreement between model predictions and expert review on the image level and patient level. The algorithm generated predictions at a rate of 0.91 seconds per scan on the GPU. Compared to AI, the available metadata yielded AUC: 0.185 and AUC: 0.598 for IV contrast detection in internal validation set and external validation set, respectively. The Confusion matrices shows significant discordance between scan metadata and expert clinicians' annotations for internal validation set (Fig. S2A) and external validation set (Fig. S2B).

**Model fine tuning and validation on chest CT scans**

With model fine-tuning, EfficientNetB4 model demonstrated AUC: 1 (95% CI: 1.0-1.0), sensitivity: 100% (95% CI: 100%-100%), specificity: 100% (95% CI: 100%-100%) and $F_1$-score: 1.0 on the patient level for the internal validation set at the patient-level (Table 1; Fig. 3C). Additionally, for external validation set, the fine-tuned model yielded AUC: 0.980 (95% CI: 0.980-0.981), sensitivity: 96.9% (95% CI: 96.8%-97.0%), specificity: 95.9% (95% CI: 95.8%-96.1%) and $F_1$-score: 0.923 at the patient-level (Table 3; Fig 3D). The EfficientNetB4 confusion matrices showed 0 (out of 53 patients) incorrect prediction in the internal validation set (Fig. 4C) and 15 (out of 402 patients) incorrect predictions in the external validation set (Fig. 4D).

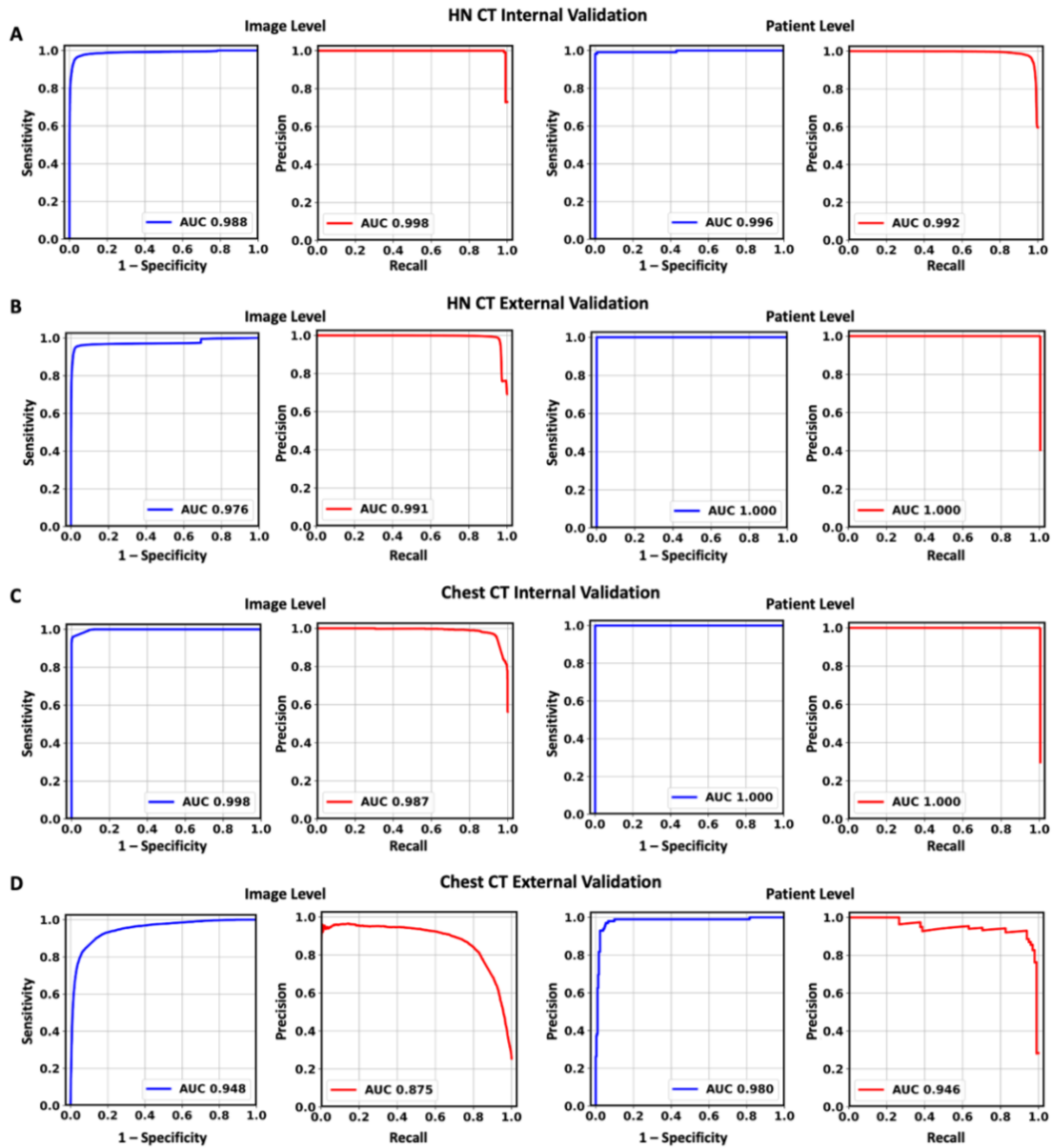

**Figure 3.** Receiver operating characteristics (ROC) curves and precision-recall (PR) curves calculated with EfficientNetB4 model on both image level and patient level for head and neck (HN) cancer internal validation set (A; n = 216 patients, 33,264 images), HN cancer external validation set (B; n = 595 patients, 39,270 images), lung cancer internal validation set (C; n = 53 patients, 3,710 images), and lung cancer external validation set (D; n = 402 patients, 28,140 images). All 6 ROC curves showed high areas under curve (AUC), indicating strong sensitivity and specificity in detecting these contrast enhancements on both image and patient levels. The PR curve of lung

CT external validation on the image level showed slightly lower AUC compared to other PR curves. HN = head and neck.

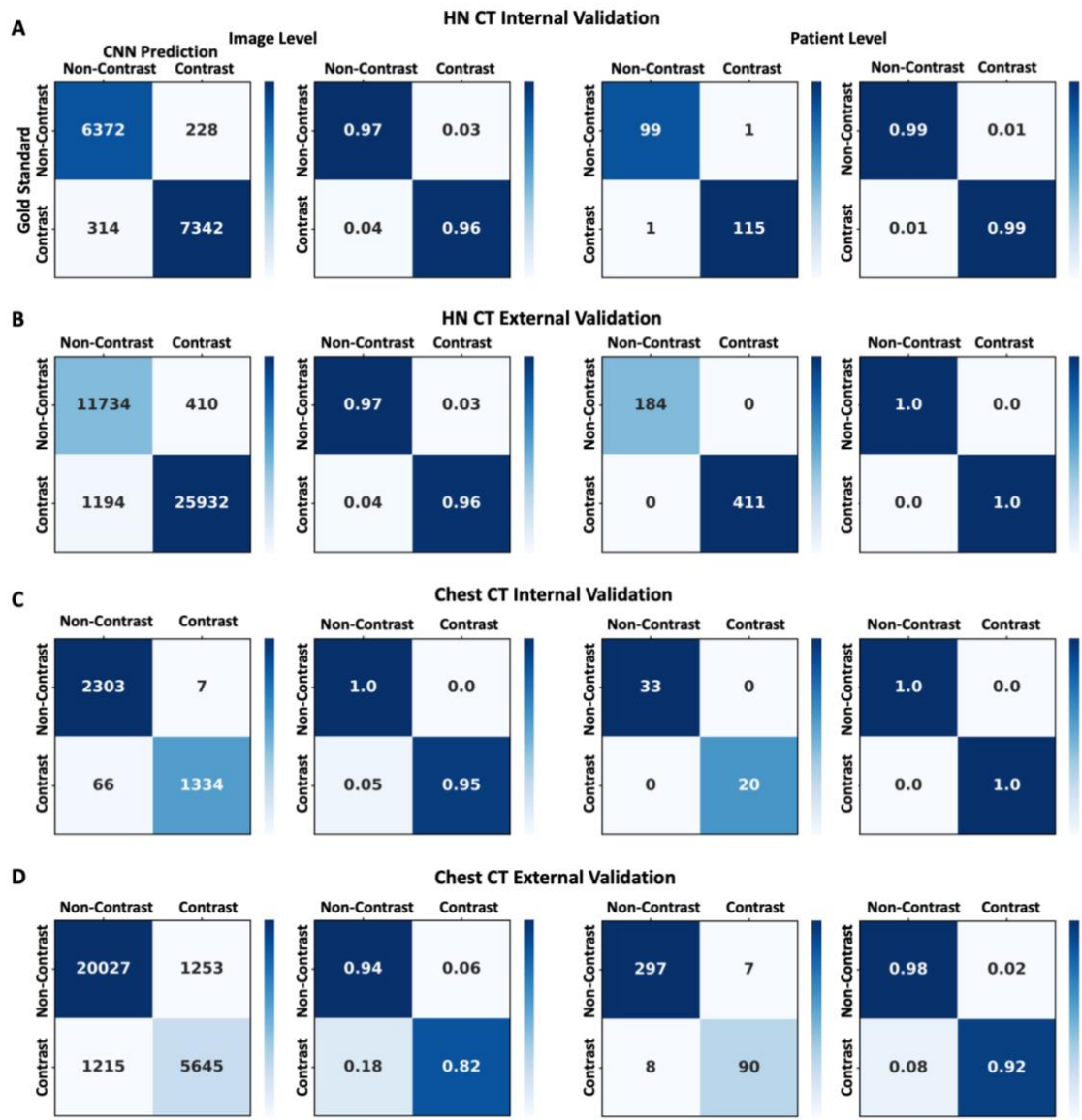

**Figures 4.** Confusion matrices calculated with EfficientNetB4 model on the internal validation data (A; n = 216 patients, 33,264 images) and external validation data (B; n = 595 patients, 39,270 images) from head and neck cancer patient cohort, as well as internal validation data (C; n = 53 patients, 3,710 images) and external validation data (D; n = 402 patients, 28,140 images) from lung cancer patient cohort. Classifications on both image level and patient level were included. Rows contain contrast enhancement classifications identified by expert physicians ("gold standard"). Columns contain lesion classifications as predicted by DNN models. HN = head and neck.

| Scan Types | Validation Types | Evaluation Levels | AUC (95% CI) | Sensitivity (%) (95% CI) | Specificity (%) (95% CI) | $F_1$-score |
|---|---|---|---|---|---|---|
| HN CT | Internal Validation | Image Level (n = 33,264) | 0.988 (0.988 – 0.988) | 95.9 (95.9 – 96.0) | 96.6 (96.6 – 96.7) | 0.964 |
| | | Patient Level (n = 216) | 0.996 (0.996 – 0.996) | 98.9 (98.8 – 98.9) | 99.9 (99.8 – 99.9) | 0.991 |
| | External Validation | Image Level (n = 39,270) | 0.976 (0.976 – 0.976) | 95.1 (95.1 – 95.1) | 97.5 (97.5 – 97.6) | 0.970 |
| | | Patient Level (n = 595) | 1 (1 – 1) | 100 (100 – 100) | 100 (100 -100) | 0.999 |
| Lung CT | Internal Validation | Image Level (n = 3,710) | 0.998 (0.998 – 0.998) | 96.0 (95.9 – 96.0) | 99.6 (99.5 – 99.6) | 0.973 |
| | | Patient Level (n = 53) | 1 (1 – 1) | 100 (100 – 100) | 100 (100 – 100) | 1 |
| | External Validation | Image Level (n = 28140) | 0.948 (0.948 – 0.949) | 85.8 (85.6 – 85.9) | 91.4 (91.3 – 91.5) | 0.821 |
| | | Patient Level (n = 402) | 0.980 (0.980 – 0.981) | 96.9 (96.8 – 97.0) | 95.9 (95.8 – 96.1) | 0.923 |

**Table 3.** Classification performances of EfficientNetB4 model on head & neck (HN) CT scans and lung CT scans. CI = confidence interval.

**Analysis of failures**

The EfficientNetB4 model had very few patient-level failures on any validation set. On manual review of the failures for the internal validation head and neck CT dataset (n = 2), we found that both cases have severe metal streak image artifacts. However, all other scans with artifact (n = 141) were still correctly classified in this internal validation set. On manual review of the failures for chest CT (n = 15), we found that 3 scans had faint contrast and were incorrectly predicted to have no contrast. The probability scores for the incorrectly classified cases were generally close to the 0.5 threshold (mean score±standard deviation: 0.55 ± 0.26).

**Gradient-weighted class activation maps for model interpretability**

Grad-CAMs were generated with EfficientNetB4 model to localize the anatomy that contributed to predictions. We extracted Grad-CAM heatmaps for the last convolutional layers of our model and selected 3 representative cases each from both HN CT dataset and chest CT dataset (Fig. 5). This provided a spatial representation of areas within the input images that contribute the most to the model prediction. Qualitative analysis of heatmaps demonstrated that regions of importance are clustered around the central blood vessels of the neck and chest, fitting the biological hypothesis.

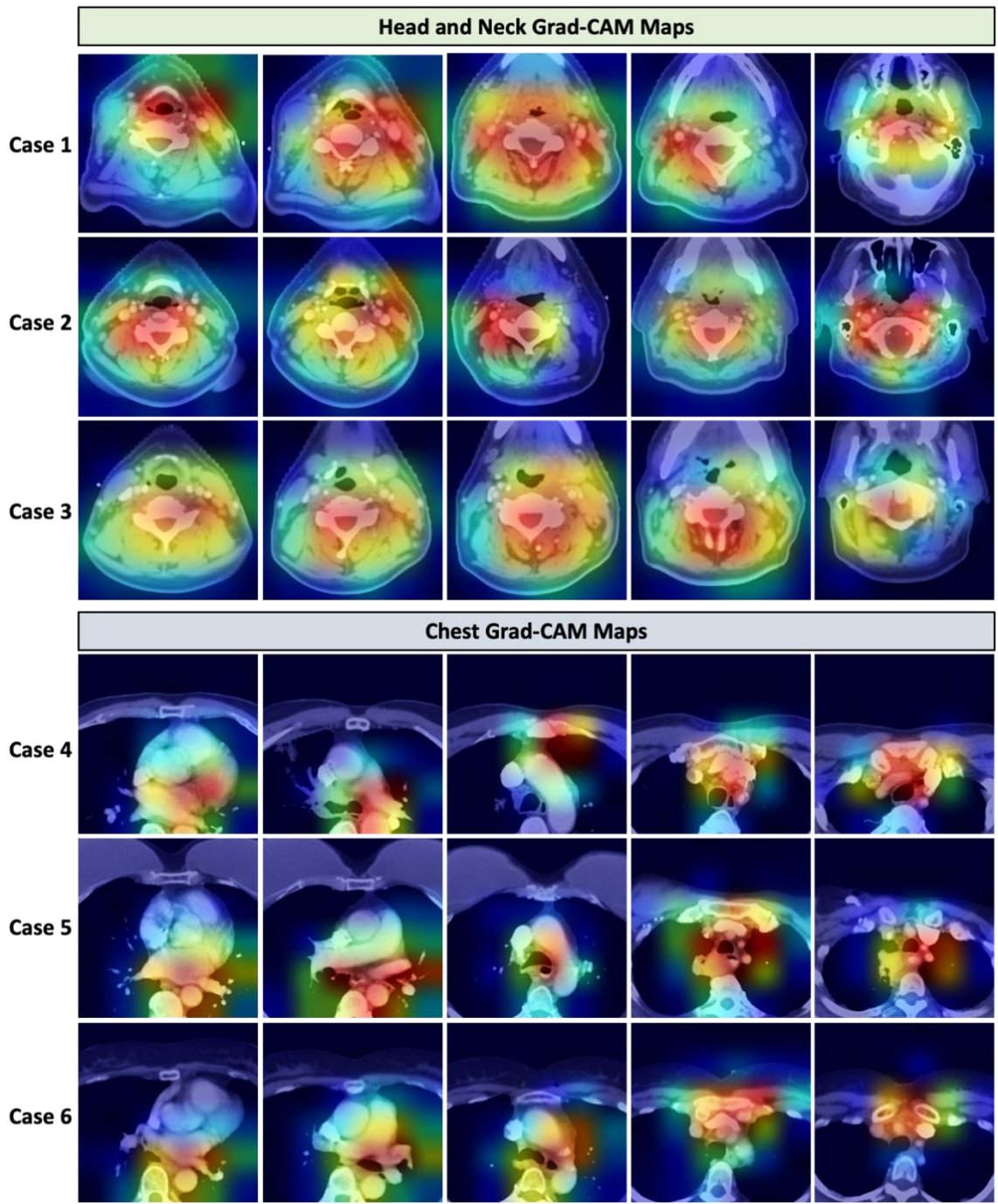

**Figures 5.** Gradient based class activation maps (Grad-CAM) from EfficientNetB4 model. Six representative scans from head and neck cancer patients (Cases 1-3) and lung cancer patients (cases 4-6) with five different image slices were shown. The last convolutional layer in model were used for generation of class activation maps. Test input images are shown with overlaid activation

maps, where red colors highlight regions with higher contribution and blue representing areas with the lower weight value.

**DISCUSSION**

We developed a CNN-based deep learning platform for automated IV contrast detection in CT scans and demonstrated that this strategy could result in near-perfect classification performance on several large datasets from a variety of institutions, treatment, and scanner settings. We further demonstrated that, with fine-tuning on additional small datasets, an IV contrast model developed for one anatomic site (head and neck), could be successfully applied to another region (chest). We have made this pipeline publicly available for further tests and applications.

Data curation and quality assessment are extremely time-limiting, resource-intensive, and manual aspects of AI algorithm development currently, and the task of annotating IV contrast for CT imaging data is no exception. Similar to prior studies (7–9), we found that IV contrast annotation from DICOM clinical metadata is often poorly documented and unreliable, with over 70% of scans in our study missing or containing erroneous contrast status, perhaps due to reliance on manual entry and institutional variation in documenting this metadata. Given the impact of IV contrast on the results of imaging analysis studies (11–14), manual review and annotation are recommended, though these are time and labor-intensive and require medical experts. Our model pipeline can efficiently capture large volumes of CT scans in DICOM format and generate near-perfect IV contrast annotations quickly (AI vs. clinician: 0.9 second/scan vs. 20 seconds/scan), providing a useful tool for the clinical and research imaging analysis communities that may obviate the need for manual annotation and review.

This is the first deep learning-based application automated CT contrast detection in the medical field, and the level of performance suggests it is suitable for real-world use. There have been several computer vision models developed for CT contrast detection or contrast phase classification that have demonstrated adequate performances. Criminisi et al (2011) proposed a hybrid discriminative-generative probabilistic model to detect CT IV contrast for multiple organs and showed a detection accuracy of >91% at ~1 second per scan on a dataset of 400 patients CT scans (8). In a study similar to IV contrast detection, Sofka et al (2011) used multi-class LogitBoost classifiers to characterize CT contrast phases (native, hepatic dominant, hepatic venous and equilibrium) with an accuracy of 93% on a test set of ~300 scans on an average speed of 7 seconds per scan (9). Recently, Tang et al (2020) proposed an unpaired contrast disentangling generative adversarial network (CD-GAN) discriminator follows the ResNet architecture to classify CT contrast phases (non-contrast, portal venous and delayed) (34) and demonstrated an accuracy of 0.91 on 9100 slices from 30 independent subjects. Compared to these previous studies, we demonstrated exceptional contrast detection ability of our model with a more comprehensive deep learning study and much larger datasets from multiple institutions with external validations. Notably, our activation map studies show neck and mediastinal blood vessels as regions of network attention in predicting contrast, supporting that the model predictions are based on robust biological phenomena. Additionally, our analysis of failure cases demonstrated very few issues overall, mostly around metal imaging artifacts. We further demonstrate the robustness of our model by using CT data with different imaging protocols from various scanner models and vendors, across multiple institutions, slice thicknesses and spatial resolutions.

There are several limitations of this study. Firstly, the study was limited to investigation of the four described 2D CNN architectures, and we did not explore 3D CNN models. The latter are widely known to be difficult to train and would likely be computationally prohibitive for use in the setting of whole-scan analysis (35). Additionally, it is more difficult to leverage transfer learning and fine-tuning in the 3D setting. Secondly, there may be uncaptured confounders within our datasets that vary from data obtained at other institutions. We recommend future users of the pipeline to conduct small local tests on their institutional scans prior to implementation at scale. Lastly, while we expect the model to perform well in other anatomic regions with further fine-tuning (e.g. the abdomen and pelvis), this was outside the scope of the current study. In future work, we will extrapolate this model to other parts of the body, and we encourage other groups to investigate our pipeline in various settings to further explore this question.

In conclusion, we demonstrated that a CNN-based deep learning model accurately detects IV contrast enhancement in CT scans from both head and neck and lung cancer patients across multiple institutions with efficiency and near perfect performance, enabling scan-to-prediction automated contrast detection. We have made publicly available this pipeline to assist data curation, quality control, and model deployment for a wide range of clinical and research applications.

**Acknowledgement**

The authors acknowledge support from the National Institutes of Health (NIH) with grant numbers (NIH U24CA194354, NIH U01CA190234, NIH U01CA209414, NIH R35CA22052, NIH K08DE030216), the European Union-European Research Council (866504), and the Radiological Society of North America. D.P. was supported by NIGMS grant T32-GM007753. This manuscript was prepared using data from Datasets [RTOG-0617] from the NCTN Data Archive of the National Cancer Institute's (NCI's) National Clinical Trials Network (NCTN). Data were originally collected from clinical trial NCT number [NCT00533949] [High-Dose or Standard-Dose Radiation Therapy and Chemotherapy With or Without Cetuximab in Treating Patients With Newly Diagnosed Stage III Non-Small Cell Lung Cancer That Cannot Be Removed by Surgery]. All analyses and conclusions in this manuscript are the sole responsibility of the authors and do not necessarily reflect the opinions or views of the clinical trial investigators, the NCTN, or the NCI.

**Competing interests**

The other authors declare no conflict of interests.

**Author contributions**

Study design: B.H.K., Z.Y., R.M., H.J.W.L.A.; code design, implementation and execution: Z.Y., B.H.K.; acquisition, analysis or interpretation of data: Z.Y., B.H.K., J.Q., A.H., R.Z., D.P., J.L., Z.Z.; image annotation: J.Q., B.H.K.; writing of the manuscript: Z.Y, B.H.K., R.M., H.J.W.L.A.; critical revision of the manuscript for important intellectual content: all authors; statistical analysis: Z.Y., B.H.K.; study supervision: B.H.K., R.M., H.J.W.L.A.

**Data availability**

NLST data including raw CT images may be requested from The Cancer Image Archive (https://www.cancerimagingarchive.net). Although raw CT imaging data cannot be shared, all measured results to replicate the statistical analysis are shared at the Github webpage at https://github.com/xmuyzz/DeepContrast. Furthermore, we include test samples from a publicly available data set with deep learning and expert reader annotations.

**Code availability**

The code of the deep learning system, as well as the trained model and statistical analysis are publicly available at the Github webpage: https://github.com/xmuyzz/DeepContrast. A working model of the EfficientNet contrast predictor can be found on Modelhub.ai.

**SUPPLEMENTARY MATERIALS**

**SUPPLEMENTARY METHODS**

**Imaging protocols**

All the scan information was summarized in Tables S2, 3&4. CT scanner model, slice thickness, pixel spacing, scanner energy, reconstruction. Table S1. CT imaging parameters used in the study. Images in the head and neck CT dataset were acquired using either of: GE LightSpeed Plus (n = 91), GE Discovery ST (n = 44), GE (unknown model, n = 286), Toshiba Aquilion One (n = 160), Toshiba Aquilion (n = 8), Toshiba (unknown model, n = 5), Philips GeminiGXL 16 (n = 101), Philips Brilliance Big Bore (n = 3), Philips (unknown model, n = 226), TomoTherapy Incorporated Hi-Art (n = 61). Imaging parameters of the other publicly available datasets have been described in the corresponding publications.

**Study patient datasets**

The head and neck (HN) cancer dataset consists of four publicly available, de-identified patient cohorts, each downloaded and curated from The Cancer Imaging Archive (TCIA) (1), as follows: 1) PMH cohort (n = 558): subset of radiotherapy planning CT scans retrieved for all non-metastatic p16-positive Oropharynx cancer (OPC) patients treated with radiotherapy or chemoradiotherapy at Princess Margaret Cancer Centre between 2005 and 2010 (2); 2) CHUM cohort (n = 61) & CHUS cohort (n = 101): subsets from FDG-PET/CT and radiotherapy planning CT imaging data of 298 head and neck cancer patients from four different institutions in Québec between 2006 and 2014 (3); 3) MDACC cohort (n = 603): subset from 627 head and neck squamous cell carcinoma (HNSCC) patients treated at MD Anderson Cancer Center (4). The lung cancer dataset includes two patient cohorts: 1) Harvard-RT cohort (n = 262): subset CT scans from non-small cell lung cancer (NSCLC) patients treated with radiation therapy at Dana-Farber Cancer Institute from 2001 through 2015; 2) RTOG-0617 cohort (n = 402): subset of the study of high-dose or standard-dose radiation therapy and chemotherapy with or without cetuximab in treating patients with newly diagnosed stage III NSCLC that cannot be removed by surgery (5). Data from PMH, CHUM and CHUS cohorts was used for the training and interval validation for the HN model. Data from MDACC cohort was used for the external validation for the HN model. Additionally, data from the Harvard-RT cohort data was used to fine tune and internally validate the HN model for chest CT. Data from the RTOG-0617 cohort was used for the external validation for the CT chest fine-tuned model (Fig. 1).

**CT image cropping**

After image co-registration, we applied a cropping step on three dimensions for each of the scans. Specifically, the original 512×512 mm$^2$ size in x-y plane was cropped to 192×192 mm$^2$ to fit the approximate axial plane boundaries of the human head and neck. To reduce memory usage and computational load, we cropped the image in the z direction and took two thirds of slices from the center to cover the entire neck and lower portion of the head (Fig. 2A). We used the same cropping step for lung cancer data to cover the center portion of the chest, where most major blood vessels are located (Fig. 2A).

**Deep convolutional neural network architectures**

In brief, our simple CNN model was constructed with 4 convolutional layers and 2 fully connected layers. Rectified linear units (ReLU) were used to activate specific functions in each hidden layer. Batch normalization was performed before feeding data to the next hidden layer to improve model optimization. Dropout layers with a rate of 0.3 were employed to prevent overfitting. The final layer was a fully connected sigmoid layer that produces probability for the positive class. For ResNet101V2, InceptionV3 and EfficientNetB4 models, the last fully-connected layers were replaced with one fully-connected layer and a sigmoid layer. In transfer learning approach based on ResNet101V2 with pre-training weights on ImageNet, we froze all layers in the base model and only trained the first fully-connected layers. As soon the model reached convergence, we trained the base model and fine-tuned the entire model end-to-end with a low learning rate ($10^{-5}$).

**CNN model implementation, training and validation**

We first trained, tuned, and tested the model on the image level. After image preprocessing, we generated 33,264, 14,256, and 39,270 images for the training, internal validation, and external validation sets, respectively. All four of the aforementioned models were independently trained. During training, the Adam optimizer was used with a mini-batch size of 32 and initial learning rate of $10^{-5}$. The learning rate was manually tuned to achieve the fastest convergence. We used binary cross-entropy loss function and trained the model to minimize the error rate of IV contrast presence on the validation dataset. Each CNN model was trained for up to 100 epochs on the training dataset and validated after each epoch on the validation set. Validation loss was calculated after each epoch and model weights were saved following each epoch that showed improvement in validation loss. Training was stopped early if 20 consecutive epochs passed without improvement in the validation loss. The hyper-parameters and optimization algorithms were chosen through a combination of grid search and manual tuning. To avoid overfitting, we utilized data augmentation on the training set with a series of random rotations, as well as flipping and zooming steps. Specifically, we used 5 degrees of random rotations and a 0.1 zoom factor.

**SUPPLEMENTARY TABLES**

| Artifacts | Training (n = 504) | Internal Validation (n = 216) | External Validation (n = 595) |
|---|---|---|---|
| No | 165 (32.7%) | 65 (30.1%) | NA |
| Minimal | 30 (6.0%) | 9 (4.2%) | NA |
| Moderate | 129 (25.6%) | 81 (37.5%) | NA |
| Severe | 149 (29.6%) | 61 (23.6%) | NA |
| Very Severe | 31 (6.2%) | 10 (4.6%) | NA |

**Table S1.** CT scan artifacts for head and neck scans (n = 1,315 patient scans).

| Manufacturer | Model | Patients n (%) |
|---|---|---|
| GE Medical Systems | Discovery ST | 312 (23.7%) |
| GE Medical Systems | Unspecified | 269 (20.5%) |

| GE Medical Systems | LightSpeed Plus | 87 (6.6%) |
| --- | --- | --- |
| Toshiba | Aquilion ONE | 149 (11.3%) |
| Toshiba | Aquilion | 7 (0.5%) |
| Toshiba | Unspecified | 5 (0.4%) |
| Philips | Unspecified | 117 (8.9%) |
| Philips | GeminiGXL 16 | 101 (7.7%) |
| Philips | Brilliance Big Bore | 3 (0.2%) |
| TomoTherapy Incorporated | Hi-Art | 61 (4.6%) |
| Unspecified | Unspecified | 29 (2.2%) |
|  | Total | 1,315 (100%) |

**Table S2.** CT scanner manufacturers and models used for head and neck scans (n = 1,315 patient scans).

| Manufacturer | Model | Patients n (%) |
| --- | --- | --- |
| GE Medical Systems | LightSpeed RT | 145 (21.8%) |
| GE Medical Systems | LightSpeed RT 16 | 98 (14.8%) |
| GE Medical Systems | LightSpeed QX/i | 14 (2.1%) |
| GE Medical Systems | Discovery CT590 RT | 1 (0.2%) |
| Unspecified | Unspecified | 402 (60.5%) |
|  | Total | 664 (100%) |

**Table S3.** CT scanner manufacturers and models used for chest scans (n = 664 patient scans).

| Scan Characteristic | Mean | Median | Mode | Range (min – max) | Standard Deviation |
| --- | --- | --- | --- | --- | --- |
| Pixel Size (cm) | 0.89 | 0.98 | 0.98 | (0.34 – 2.34) | 0.33 |
| Slice Thickness (cm) | 2.2 | 2.5 | 2.5 | (1.0 – 5.0) | 0.70 |
| Tube Voltage (kVp) | 125.0 | 120.0 | 120.0 | (120.0 – 140.0) | 8.39 |

**Table S4.** Head and neck CT scan characteristic deviation table (n = 1,296 patient scans).

| Scan Characteristic | Mean | Median | Mode | Range (min – max) | Standard Deviation |
|---|---|---|---|---|---|
| Pixel Size (cm) | 1.07 | 0.98 | 0.98 | (0.65 – 2.73) | 0.24 |
| Slice Thickness (cm) | 2.7 | 2.5 | 2.5 | (1.0 – 5.0) | 0.38 |
| Tube Voltage (kVp) | 120.2 | 120.0 | 120.0 | (120.0 – 140.0) | 1.76 |

**Table S5.** Chest CT scan characteristic deviation table (n = 632 patient scans). Data was only available for 632 patients.

| Slice Thickness (cm) | n (%) |
|---|---|
| 5.00 | 8 (0.6%) |
| 3.75 | 13 (1.0 %) |
| 3.00 | 276 (21.0%) |
| 2.50 | 481 (36.6%) |
| 2.00 | 198 (15.1%) |
| 1.50 | 61 (4.6%) |
| 1.25 | 366 (27.8%) |
| 1.00 | 1 (0.08%) |
| Unspecified | 29 (2.2%) |
| Total | 1315 (100%) |
| **Axial Spatial Resolution (pixels)** | **n (%)** |
| 512 x 512 | 1235 (93.9%) |
| 272 x 272 | 51 (3.9%) |
| 256 x 256 | 10 (0.8%) |
| Unspecified | 29 (2.2%) |
| Total | 1315 (100%) |

**Table S6.** CT scan characteristic distribution table (n = 1315 patient scans).

| Slice Thickness (cm) | n (%) |
|---|---|
| 5.00 | 6 (0.9%) |
| 4.00 | 1 (0.2%) |
| 3.75 | 10 (1.5%) |
| 3.20 | 1 (0.2%) |
| 3.00 | 197 (29.7%) |
| 2.50 | 396 (59.6%) |
| 2.00 | 15 (2.3%) |
| 1.50 | 1 (0.2%) |
| Unspecified | 29 (4.4%) |
| Total | 664 (100%) |
| **Axial Spatial Resolution (pixels)** | **n (%)** |
| 768 x 768 | 1 (0.2%) |
| 512 x 512 | 609 (91.7%) |
| 272 x 272 | 4 (0.6%) |
| 256 x 256 | 18 (2.7%) |
| Unspecified | 29 (4.4%) |
| Total | 664 (100%) |

**Table S7.** Chest CT scan characteristic distribution table (n = 664 patient scans).

| **Contrast Bolus Agent Information** |
|---|
| nan, '100ML OMNI 300', '120CC OPTI', 'IV VISIPAQUE', 'OPTIRAY 350', 'VISIPAQUE IV', '125CC/2.5R/90SEC', '90CC/2.0R/80SEC', 'OMNI', 'IV', '120CC OR', 'N', 'ISOVIEW', 'ORAL/RECTAL & IV', '120CC/3.0R/90SEC', '120CC', 'OPTI 320', '123CC/3.0R/90SEC', 'OMNI 350 75ML', '125MLOPTIRAY320', 'iv', 'ISOVUE 300 150ML', '125CCOPTI', '110CC/3.0R/90SEC', '121CC/3.0R/90SEC', 'ISOVUE 370 100CC', '125CC', 'OP', 'ISOVUE 370', 'OPTI', 'OPT', 'NO', '100ML OMNI 350', 'ISOVUE', 'OMNIPAQUE 300', '125CC/3.0R/90SEC', 'opti', '100 ML ISOVUE', '3CC OPTI', '& IV', 'VISIPAQUE', '69CC', 'OPTIRAY', 'ISO 370', 'CE', 'OMNI 350 100CC', '90CC/2.0R/90SEC', 'OPTI 350', '100 CC ISOVUE 300', 'OMNIPAQUE', '115CC/3.0R/90SEC', 'IV CONTRAST', 'CONT', 'n' |

**Table S8.** Information on contrast bolus information.

|  | Internal Validation | | | | External Validation | | | |
|---|---|---|---|---|---|---|---|---|
| **DNN Models** | **Image Level (n = 33,264)** | | **Patient Level (n = 216)** | | **Image Level (n = 39,270)** | | **Patient Level (n = 595)** | |
|  | AUC | $F_1$-score | AUC | $F_1$-score | AUC | $F_1$-score | AUC | $F_1$-score |
| Simple CNN | 0.936 | 0.814 | 0.998 | 0.853 | 0.952 | 0.857 | 0.953 | 0.870 |
| ResNet101V2 | 0.989 | 0.959 | 0.993 | 0.991 | 0.987 | 0.964 | 1.0 | 1.0 |
| Transfer Learning | 0.903 | 0.790 | 1.0 | 0.960 | 0.928 | 0.879 | 1.0 | 0.986 |
| InceptionV3 | 0.993 | 0.968 | 0.996 | 0.987 | 0.973 | 0.970 | 1.0 | 0.999 |
| EfficientNetB4 | 0.988 | 0.964 | 0.996 | 0.991 | 0.976 | 0.970 | 1.0 | 1.0 |

**Table S9.** Diagnostic performances of different DNN models on head and neck cancer datasets.

|  | Internal Validation | | | | External Validation | | | |
|---|---|---|---|---|---|---|---|---|
| **DNN Models** | **Image Level (n = 3,710)** | | **Patient Level (n = 53)** | | **Image Level (n = 28,140)** | | **Patient Level (n = 402)** | |
|  | AUC | $F_1$-score | AUC | $F_1$-score | AUC | $F_1$-score | AUC | $F_1$-score |
| Simple CNN | 0.923 | 0.709 | 0.998 | 0.833 | 0.852 | 0.551 | 0.964 | 0.683 |
| ResNet101V2 | 0.985 | 0.933 | 1.0 | 1.0 | 0.941 | 0.806 | 0.969 | 0.904 |
| Transfer Learning | 0.975 | 0.902 | 1.0 | 1.0 | 0.950 | 0.823 | 0.984 | 0.918 |
| InceptionV3 | 0.985 | 0.916 | 1.0 | 1.0 | 0.926 | 0.787 | 0.959 | 0.880 |
| EfficientNetB4 | 0.998 | 0.973 | 1.0 | 1.0 | 0.948 | 0.821 | 0.980 | 0.923 |

**Table S10.** Diagnostic performance of different DNN models on lung cancer datasets.

| Scan Types | Validation Types | Evaluation Levels | AUC (95% CI) | Sensitivity (%) (95% CI) | Specificity (%) (95% CI) | $F_1$-score |
|---|---|---|---|---|---|---|

| Scan Types | Validation Types | Evaluation Levels | AUC (95% CI) | Sensitivity (%) (95% CI) | Specificity (%) (95% CI) | F$_1$-score |
|---|---|---|---|---|---|---|
| **HN CT** | Internal Validation | Image Level (n = 33,264) | 0.936 (0.936 – 0.936) | 85.6 (85.6 - 85.6) | 86.0 (85.9 – 86.1) | 0.814 |
| | | Patient Level (n = 216) | 0.998 (0.998 – 0.998) | 98.9 (98.8 – 99.0) | 97.5 (97.4 – 97.6) | 0.853 |
| | External Validation | Image Level (n = 39,270) | 0.952 (0.952 – 0.952) | 95.1 (95.1 – 95.1) | 97.5 (97.5 – 97.6) | 0.970 |
| | | Patient Level (n = 595) | 0.953 (0.952 – 0.953) | 100 (100 – 100) | 100 (100 -100) | 0.999 |
| **Lung CT** | Internal Validation | Image Level (n = 3,710) | 0.923 (0.923 – 0.924) | 77.7 (77.6 – 77.8) | 92.5 (925 – 92.6) | 0.709 |
| | | Patient Level (n = 53) | 0.998 (0.998 – 0.999) | 99.4 (99.3 – 99.5) | 98.8 (98.7– 99.0) | 0.833 |
| | External Validation | Image Level (n = 28,140) | 0.852 (0.852 – 0.852) | 69.8 (69.7 – 69.9) | 83.6 (83.5– 83.7) | 0.551 |
| | | Patient Level (n = 402) | 0.962 (0.961– 0.963) | 97.0 (96.9 – 97.1) | 92.6 (92.5 – 92.8) | 0.685 |

**Table S11.** Diagnostic performance of different simple CNN models. CI = confidence interval.

| Scan Types | Validation Types | Evaluation Levels | AUC (95% CI) | Sensitivity (%) (95% CI) | Specificity (%) (95% CI) | F$_1$-score |
|---|---|---|---|---|---|---|
| **HN CT** | Internal Validation | Image Level (n = 33,264) | 0.989 (0.989 – 0.989) | 95.2 (95.2 – 95.3) | 96.4 (96.4 – 96.5) | 0.959 |
| | | Patient Level (n = 216) | 0.993 (0.993 – 0.994) | 98.8 (98.8 – 98.9) | 99.8 (99.8 – 99.8) | 0.991 |
| | External Validation | Image Level (n = 39,270) | 0.973 (0.973 – 0.973) | 94.8 (94.8 – 94.9) | 98.1 (98.1 – 98.1) | 0.970 |
| | | Patient Level (n = 595) | 1.0 (1.0 – 1.0) | 100 (100 – 100) | 100 (100 – 100) | 0.999 |
| **Lung CT** | Internal Validation | Image Level (n = 3,710) | 0.998 (0.9298 – 0.998) | 95.9 (95.9 – 96.0) | 99.6 (99.5 – 99.6) | 0.973 |
| | | Patient Level (n = 53) | 1.0 (1.0 – 1.0) | 100 (100 – 100) | 100 (100 – 100) | 1.0 |
| | External Validation | Image Level (n = 28,140) | 0.941 (0.925 – 0.925) | 86.8 (86.6 – 87.0) | 88.1 (87.9 – 88.4) | 0.806 |

| | | Patient Level (n = 402) | 0.969 (0.968 – 0.970) | 95.3 (95.2– 95.5) | 95.9 (95.8 – 95.9) | 0.904 |

**Table S12.** Diagnostic performance of different simple ResNet101V2 model. CI = confidence interval.

| Scan Types | Validation Types | Evaluation Levels | AUC (95% CI) | Sensitivity (%) (95% CI) | Specificity (%) (95% CI) | $F_1$-score |
|---|---|---|---|---|---|---|
| HN CT | Internal Validation | Image Level (n = 33,264) | 0.903 (0.903 – 0.903) | 83.5 (83.4– 83.6) | 80.4 (80.3 – 80.5) | 0.790 |
| | | Patient Level (n = 216) | 1.0 (1.0 – 1.0) | 99.2 (99.1 – 99.2) | 99.7 (99.7 – 99.7) | 0.950 |
| | External Validation | Image Level (n = 39,270) | 0.928 (0.928 – 0.928) | 83.4 (83.3 – 83.5) | 87.3 (87.2 – 87.4) | 0.879 |
| | | Patient Level (n = 595) | 1.0 (1.0 – 1.0) | 99.3 (99.3 – 99.3) | 99.9 (99.9– 99.9) | 0.986 |
| Lung CT | Internal Validation | Image Level (n = 3,710) | 0.975 (0.975 – 0.975) | 90.8 (90.8 – 90.9) | 93.7 (93.7 – 93.7) | 0.902 |
| | | Patient Level (n = 53) | 1.0 (1.0 – 1.0) | 100 (100 – 100) | 100 (100 – 100) | 1.0 |
| | External Validation | Image Level (n = 28,140) | 0.950 (0.950 – 0.950) | 86.1 (85.9– 86.3) | 90.7 (90.5 – 90.9) | 0.823 |
| | | Patient Level (n = 402) | 0.984 (0.982 – 0.983) | 98.2 (98.2 – 98.3) | 96.4 (96.3 – 96.5) | 0.918 |

**Table S13.** Diagnostic performance of different simple transfer learning based on ResNet101 V2 model.

| Scan Types | Validation Types | Evaluation Levels | AUC (95% CI) | Sensitivity (%) (95% CI) | Specificity (%) (95% CI) | $F_1$-score |
|---|---|---|---|---|---|---|
| HN CT | Internal Validation | Image Level (n = 33,264) | 0.993 (0.993 – 0.993) | 96.7 (96.7 - 96.7) | 96.9 (96.8 – 96.9) | 0.968 |
| | | Patient Level (n = 216) | 0.996 (0.996 – 0.996) | 98.9 (98.8 – 98.9) | 99.9 (99.8 – 99.9) | 0.987 |
|  | External Validation | Image Level (n = 39,270) | 0.973 (0.973 – 0.973) | 94.8 (94.8 - 94.9) | 98.2 (98.1 – 98.2) | 0.970 |
|  |  | Patient Level (n = 595) | 1.0 (1.0 – 1.0) | 100 (100 – 100) | 100 (100 – 100) | 0.999 |
| **Lung CT** | Internal Validation | Image Level (n = 3,710) | 0.998 (0.998 – 0.998) | 96.0 (95.9 – 96.0) | 99.5 (99.5 – 99.6) | 0.973 |
|  |  | Patient Level (n = 53) | 1.0 (1.0 – 1.0) | 100 (100 – 100) | 100 (100 – 100) | 1.0 |
|  | External Validation | Image Level (n = 28,140) | 0.926 (0.926 – 0.926) | 88.4 (88.4 – 88.4) | 84.9 (84.9 – 84.9) | 0.787 |
|  |  | Patient Level (n = 402) | 0.959 (0.958 – 0.960) | 94.8 (94.6 – 94.9) | 95.8 (95.7 – 95.9) | 0.880 |

**Table S14.** Diagnostic performance of different simple InceptionV3 model.

## SUPPLEMENTARY FIGURES

```
_________________________________________________________________
Layer (type)                 Output Shape              Param #
=================================================================
conv2d (Conv2D)              (None, 190, 190, 16)      448
_________________________________________________________________
batch_normalization (BatchNo (None, 190, 190, 16)      64
_________________________________________________________________
max_pooling2d (MaxPooling2D) (None, 95, 95, 16)        0
_________________________________________________________________
dropout (Dropout)            (None, 95, 95, 16)        0
_________________________________________________________________
batch_normalization_1 (Batch (None, 95, 95, 16)        64
_________________________________________________________________
conv2d_1 (Conv2D)            (None, 93, 93, 64)        9280
_________________________________________________________________
max_pooling2d_1 (MaxPooling2 (None, 46, 46, 64)        0
_________________________________________________________________
dropout_1 (Dropout)          (None, 46, 46, 64)        0
_________________________________________________________________
batch_normalization_2 (Batch (None, 46, 46, 64)        256
_________________________________________________________________
conv2d_2 (Conv2D)            (None, 44, 44, 128)       73856
_________________________________________________________________
max_pooling2d_2 (MaxPooling2 (None, 22, 22, 128)       0
_________________________________________________________________
dropout_2 (Dropout)          (None, 22, 22, 128)       0
_________________________________________________________________
batch_normalization_3 (Batch (None, 22, 22, 128)       512
_________________________________________________________________
conv2d_3 (Conv2D)            (None, 20, 20, 128)       147584
_________________________________________________________________
max_pooling2d_3 (MaxPooling2 (None, 10, 10, 128)       0
_________________________________________________________________
dropout_3 (Dropout)          (None, 10, 10, 128)       0
_________________________________________________________________
flatten (Flatten)            (None, 12800)             0
_________________________________________________________________
batch_normalization_4 (Batch (None, 12800)             51200
_________________________________________________________________
dense (Dense)                (None, 256)               3277056
_________________________________________________________________
dropout_4 (Dropout)          (None, 256)               0
_________________________________________________________________
batch_normalization_5 (Batch (None, 256)               1024
_________________________________________________________________
dense_1 (Dense)              (None, 256)               65792
_________________________________________________________________
dropout_5 (Dropout)          (None, 256)               0
_________________________________________________________________
dense_2 (Dense)              (None, 1)                 257
=================================================================
```

**Figure S1.** Model architecture of simple convolutional neural network.

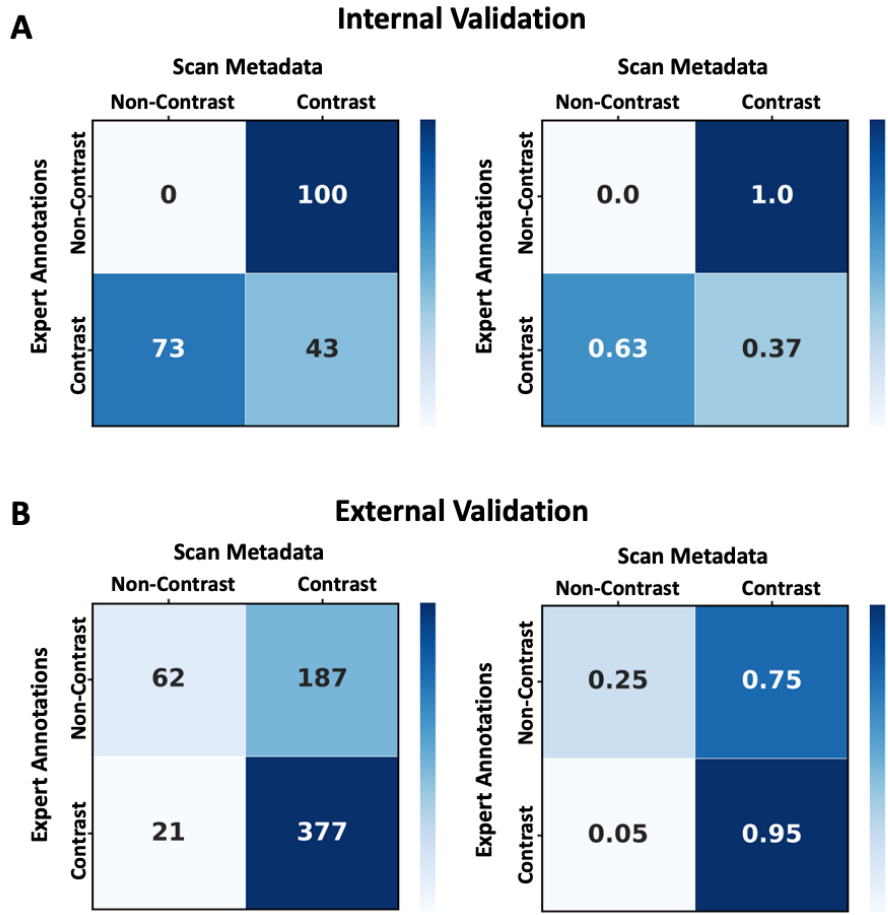

**Figure S2.** Confusion matrices showing discordance of IV contrast information between scan metadata and expert clinicians' annotations for internal validation set (A) and external validation set.